\documentclass[12pt]{article}
\def\be{\begin{equation}} \def\ee{\end{equation}}
\def\bi{\begin{itemize}} \def\ei{\end{itemize}}
\def\bea{\begin{eqnarray}} \def\eea{\end{eqnarray}} \def\ba{\begin{array}}
\def\ea{\end{array}} \def\ben{\begin{enumerate}} \def\een{\end{enumerate}}

\newcommand{\eqn}[1]{(\ref{#1})}

\newcommand{\hepth}[1]{{\tt [arXiv:{#1} [hep-th]]}}

\def\br{\nonumber\\}

\def\tr{{\rm Tr}}

\begin{document}
{}~

\hfill\vbox{\hbox{\today}}\break

\vskip 2.5cm
\centerline{\Large \bf Instantonic string solitons on M5 branes }
\vskip .5cm
\vspace*{1cm}
\centerline{ \sc Harvendra Singh }
\vspace*{.5cm}
\centerline{ \it  Theory Division, Saha Institute of Nuclear Physics} 
\centerline{ \it  1/AF Bidhannagar, Kolkata 700064, India}
\vspace{.25cm}
\centerline{ \it  Homi Bhabha National Institute, 
Anushakti Nagar, Mumbai 400094, India}
\vspace*{.25cm}

\vskip1.5cm

\centerline{\bf Abstract} \bigskip
The (I)nstantonic strings  are described as extended   states 
occupying a flat (isometry) direction on the  multiple M5-brane
world-volume. These are constituted by 4-dimensional gauge 
instantons and  2-dimensional chiral axion. In this light
we  revisit the covariant action for M5-branes  
by adding explicit axion terms. This leads to 
a new gauge symmetry
and  simple modification of some field equations and the 
 constraints  in the  theory. 
The equations of motion include a conserved toplological current of
 I-strings. 
When 6D theory is compactified on a circle the light-like
 I-strings  manifest as heavy monopole states in 5D 
super-Yang-Mills.  
While extremely light I-strings tend to become
degenerate with the M5 vacuum but only at strong SYM coupling.
This indicates that 6D vacuum is infinitely degenerate. 

\vfill 
\eject

\baselineskip=16.2pt


\section{Introduction}

The  holographic constructions of 
M2-brane theories like BLG and ABJM
\cite{blg}-\cite{hs}, 
 provided us with some much needed insight  to construct    
 similar 6-dimensional theory for multiple M5-branes. 
Some breakthrough ideas in constructing M5  theories
with maximal supersymmetry were proposed in the works
 \cite{lp,doug10,lps,hs11}, also see \cite{Basu:2004ed}.  
 Subsequently these led to  6-dimensional generalizations 
of  theories with lesser supersymmetry
   \cite{Samtleben:2011}-\cite{Lavau:2014iva}. 
The self-dual antisymmetric
 tensor fields are  natural to occur in six spacetime dimensions and
the  
dynamics of a single M5-brane can be described by an abelian tensor theory
with  (2,0) supersymmetry. 
There are similar important dynamical
reasons to include tensor fields in the 6D constructions. 
Let us consider the example of an extended M2-brane ending on 
 M5-brane. The intersection of  M2 and M5 will
produce an infinitely long line defect (string) on  six-dimensional
world-volume of  M5-brane. These  strings
perhaps constitute  simplest excitations which could
 entirely  live on the M5-brane and carry charges of tensor fields. 
Basically, these line defects 
 behave  like  extended charged strings  in 6D Minkowski
 spacetime but these strings  cannot  carry 
momentum along their direction of  extension. 
It is  believed  that ultimately the 
dynamics of these  charged strings  does
constitute the low energy  dynamics on the M5-brane. 
Similarly when we   consider 
 multi-brane  configurations where   $N$  coincident M5-branes
having a single M2-brane ending on them. In that situation the M2-brane will 
produce a line defect on each M5-brane in the stack. 
Thus we  have a low energy 
configuration  which has to be described by 
$N$ parallel (aligned) 
strings in six dimensions. Of course, such extended string configurations 
will spontaneously break  the Lorentz symmetry 
from $SO(1,5) \to SO(1,4)$. 
 Thus we may  conclude that
 low energy string excitations  on  M5-branes  
 leads to manifestly broken  Lorentzian symmetry. 
Indeed this has been an in-built criterion in some 
of the models \cite{lp,hs11}. Such an
 hypothesis allows for the inclusion of an auxiliary
vector field $\zeta_M$
in the theory. With  $\zeta_M$ and an associated axion  field, we are  
able to uplift the entire dynamics of the SYM to  six dimensions
in a  covariant manner.

\noindent{\it\bf The puzzle:} \\
The  extended strings  naturally couple to 
antisymmetric tensor field, whose field strength  
$H_{(3)}$  needs to be self-dual,
for it to describe an excitation on the M5-brane \cite{howe}. 
The self-dual tensor field along with 
{\it five} scalars, $X^I$'s, and  one Majorana-Weyl
 spinor $\Psi$, are part of the 
 (2,0) tensor supermultiplet in six  dimensions \cite{howe1}. 
The dynamical equations of this  chiral theory  are 
\bea
\label{self1a}
H_{(3)}\equiv dB_{(2)}=\star_6 H_{(3)}, ~~~
\partial_M\partial^M X^I=0=\not\!\partial\Psi
\eea
where $\star_d$ is a Hodge-dual operation in $d$-dimensional spacetime. 
However, this abelian (2,0) CFT is a free  theory. 
While a non-abelian version of the tensor theory 
is not  satisfactorily known at present. But there  is a belief  that
all  states of a non-abelian (2,0) CFT, once compactified
on a circle, are perhaps  contained in  5D 
super-Yang-Mills (SYM) theory and vice-versa. The SYM in five dimensions
is  nonrenormalizable  and has a strongly coupled fixed point in the
UV region. 
A conventional nonrenormalizable theory would require many
new degrees of freedom to be added to it at shorter and shorter  scales.  
Thus if the tree level  SYM  fields indeed describe all the states 
of a `compactified' 6D CFT, 
without requiring  new  degrees of freedom,  then 
the SYM ought to be considered  a finite theory in itself \cite{doug10,lps}. 
A test of this proposal  requires that all
Kaluza-Klein modes, if any,  
of  6D fields  are mapped into various (instantonic) monopole sectors
in the SYM. 
Although  qualitative, but it has remained a  difficult  prospect
to directly  verify  the  finiteness of the 5D SYM. 
However, one direct puzzle  to note here is that in  abelian  CFT 
there would be  no instantons, while  KK (Fourier) 
modes of the self-dual tensor field, as 
in \eqn{self1a}, are  all likely to  survive. So this identification map
of `6D KK-modes' Vs `5D monopoles' seems to be
problematic and   puzzling!
Any  deviation from
the expected/conjectured  behaviour of 5D SYM is of course
 a welcome sign as it will have 
consequences for   6D CFT.  
See for recent important 
attempts along this direction \cite{doug12,lp,lps,hs11}. 

\noindent{\it\bf Instantonic strings:}\\
In this work we  propose a simple modification of previous 6D action 
presented  in \cite{hs11,hs12}.
The  proposed  action can 
describe  new  instantonic states in the  theory which are  
string like solitonic configurations 
carrying a topological charge
\be
Q_{i-string}=\int d^5x J_{i-string}^0 =
{1\over 2}\tr
\int {\zeta}_{(1)} \wedge F_{(2)}\wedge F_{(2)}
\ee
The instantonic 
 charges  appear because of the gauge instantons which  live 
over 4D Euclidean subspace and the finite vev of 
  auxiliary vector field $\zeta_M$. Thus, in general, the 
charge  of  I-string  
is fixed by the instanton number
and will be measured in  units of $|\zeta|$ vev.
Other consequences are  that the   $S^1$ reduction produces
 { \it   no KK-modes in 5D}, because
 all six-dimensional fields have  no  dynamics along 
 isometry (circle) direction owing to the constraints in the theory. 
Instead these  I-string  states  directly
manifest as monopole solitons in 5D super-Yang-Mills theory. 

Further, a true vacuum state in field theory is the one with  6-momentum
$p_M=0$. There will also be
 solutions  which would occur in the theory that will have
broken   Poincare\'{}  symmetry. For 
the I-strings,   made up of instantons, we shall
have   
\be
p^M=(|p|,0,0,0,0,  p)
\ee
Especially the momentum entries  
directly deal with the instanton number, 
these are nonvanishing and take  discrete values.
As an example, the lightlike auxiliary vector
\be
<\zeta^M>=|\zeta|(1,0,0,0,0,1)
\ee
 does allow an I-string  instantonic state  with 
\be
p^M=( n {L |\zeta|^2 },0,0,0,0,  n {L |\zeta|^2 })
\ee
which corresponds to a lightlike state  with 
discrete energy and a fifth component of the momentum.  
Here $L$ is  regulated length (size of $x_5$ coordinate)
of the string.  Especially,
after a circle compactification of the theory
the momenta can be expressed as
\be
p^M=({n\over  g_{_{YM}}^2} ,0,0,0,0,{n\over  g_{_{YM}}^2})
\ee
These  I-string solitons are indeed 
identified with the monopoles of  5D SYM. 
We note that 
I-strings are different from other electrically (color) charged 
self-dual  string solitons \cite{howe}, which are  part of the 
spectrum but only in  {\it instanton-less} (or $n=0$) sector of the theory. 
Our main aim here is to actually show that a simple 
modification of a previous covariant 6D action  \cite{hs11} can meaningfully
represent  these  I-strings.

The paper is organised as follows. In  section-2 we introduce a new form of 
M5 action in 
six dimensions with the help of auxiliary axion field. We define a
new  conserved (instantonic) current. In section-3, we discuss 5D super 
Yang-Mills and its relationship with the 6D theory compactified on a circle. 
The  instantonic string solitons are presented 
in  section-4. It is found that while solving the equations that 
the I-string solitons are essentially  light-like states. 
These states must condense to the ground state in the decompactification limit
6D theory. We discuss  (colored) e-strings as well as  4D 
electric-magnetic solutons in the section-5. 
A brief summary is given in  section-6.    

\section{Six-dimensional tensor theories }

\subsection{  The  6D chiral  theory}

At the outset
very little is known about the full non-Abelian
(2,0) tensor theory which would  
describe the  dynamics of   M5-brane stack completely. Nontheless
meaningful attempts have been made recently 
to write down 6D  theories using self-dual tensors, at least
 at the equation of motion level \cite{lp}, 
and separately by up-lifting 5D SYM theory to the 
six-dimensions with the help of an auxiliary vector field \cite{hs11, hs12}. 
It is desirable that
 non-Abelian 6D CFT  should 
possess  $U(N)$  gauge symmetry and an $SO(5)$  R-symmetry. 
The 6D gauge action constructed in \cite{hs11} inherits
 these basic features directly from 5D  because 
it is an  uplift
of the SYM. The action  had explicit scale invariance and is devoid of 
dimensionful parameters. 
 These requirements at handy because ultimately they 
will guide us in the construction of
 M5-brane action.  
\footnote{ See more  developments on M5 
theory in the references 
\cite{witt,ps,tonin,howe2,Howe:1997fb, Lee:2000kc,tonin1}.}

It is now  known how to uplift  5D SYM to six dimensions
 such that the resulting 6D action 
has no dimensionful parameters and the 6D theory exhibits  scaling symmetry.
\footnote{The latter property will make
it  different from  ${\cal{N}}=(1,1)$ 6D YM theory, which instead
describes D5-branes. The D5-brane SYM theory has  a dimensionful 
running coupling constant. }
This has been explicitly shown at the equation of motion level for (2,0) 
supersymmetric model  with chiral fields 
\cite{lp}. A variant of  6D (2,0) theory with  covariant action 
was subsequently constructed in \cite{hs11}. 
The 6-dimensional 
gauge  action (excluding fermions) can  be written  more
symmetrically as \footnote{ The action is written in a different form 
here as compared to \cite{hs11}. Note the way the kinetic terms for scalars
appear, they also have $\zeta$ field. It does not change the dynamics 
drastically.} 
\bea\label{act3a}
S_{6D}&=&\int d^6x  \bigg[\tr\{
-{1\over 2.3! } ( \zeta_{[M}F_{NP]})^2  
 -{1\over 2.2!} (\zeta_{[N}D_{M]} X^I)^2 +{1\over 4} 
(\zeta_M)^2([X^I,X^J])^2 \}
 \br
&&
~~~~~~~ +{1\over 4!}\epsilon^{MNPQRS}
\zeta_M \partial_{N} C_{PQRS}
+{1\over 4}\tr \epsilon^{MNPQRS}
\zeta_M\partial_N \theta  F_{PQ}F_{RS} \bigg] 
\eea
where  
\be
 \zeta_{[M}F_{NP]}=\zeta_{M}F_{NP} +~ {\rm cyclic~ permutations},
\ee
 and
\be
\zeta_M\equiv \eta_M + \partial_M \theta\ ,
\ee
The action has been presented in such a way that
the auxiliary  field $\eta_M$, 
with mass dimension one, 
couples to all other fields in the action. 
Hence $\zeta_M$ (or $\eta_M$) 
 sometimes is levelled as a `dressing' field.
There is an Stueckelberg  gauge invariance
\bea
&& \delta_s \eta_M=c \partial_M \lambda(x), 
~~~~~\delta_s \theta=-c \lambda(x) \br 
&& 
 \delta_s
 C_{MNPQ}=  c \lambda(x) \tr ( F_{[MN} F_{PQ]}) \ .
\eea 
where $c$ is arbitrary constant gauge parameter.
Hence by appropriately chosing the gauge, 
$\theta(x)$  can always remain hidden 
in the whole  action except in the last term. In
actuality, the  action \eqn{act3a} differs from the action in 
 previous work \cite{hs11} only by the inclusion of this last term. They
can only be related by a  nontrivial field redefinition 
\be
 C'_{(4)}
 \equiv
 C_{(4)}+ \theta(x) \tr (F_{(2)}\wedge F_{(2)}) \ .
 \ee
The  $\theta$  is an axion 
and it  is dimensionless. The $\theta$  term will be
helpful in  describing the topological (instantonic) sector 
of the 6D theory. For example, as we shall discuss, there exist new extended
instanton configurations  where the topological charge 
is uniformally distributed along the length of an straight string. These 
instantonic strings (I-string) are  analogous to  well known
electrically  charged string (e-string)  solitons in M5 brane theory. 
We also need to make sure that action \eqn{act3a} remains finite for 
the I-strings.
We shall find that the presence of  axion field in the action
subtly modifies the dynamics  by changing the $\eta_M$ field
equation. 

In the above, the 6D Lorentzian indices  are levelled as
$M,N=(0,1,\cdots,5)$, and the internal $SO(5)$ indices are levelled as
$I,J=(6,7,8,9,10)$. While
$\mu,\nu=(0,1,\cdots,4)$ would represent 5D Minkowski indices. 
The Yang-Mills field strength is defined as
\be
F_{MN}=\partial_{[M} A_{N]} -i [A_{M}, A_{N}]
\ee
The  {\it five} scalar fields $X^I$'s
are in the adjoint representation and their covariant derivative is
defined as 
\bea
&& D_M X^I= \partial_M X^I -i[A_M,X^I] .
\eea
These are the only bosonic fields which are dynamical in  
6D theory \eqn{act3a}, the rest  are all auxiliary fields
and give rise to  constraints.
There is  manifest 
$U(N)$ gauge symmetry in the  action \eqn{act3a}  corresponding
 to the fact that it describes $N$ parallel M5 branes.
The  gauge transformations are 
\bea\label{gt1}
&&  A_{M}\to A'_{M}=U^{-1}A_{M} U - i U^{-1}\partial_{M} U\br
&&  X^I\to X'^I=U^{-1} X^I U, ~~~  
\eea
under which the action \eqn{act3a} 
remains invariant. Here $U$ is an element of
$U(N)$ gauge group.
 
Let us simplify our notations  and write 
the 2-form gauge field strength as
\bea
&&F_{(2)}\equiv D A_{(1)}= dA_{(1)} - {i\over 2} [A_{(1)}, A_{(1)}] \br
&& D X^I= d X^I - {i} [A_{(1)}, X^I] 
\eea
where  $D$ is used for covariant derivative of 
adjoint fields.  The  Bianchi identity  is  
\be 
DF_{(2)}=0\ 
\ee
The action \eqn{act3a} may  be expressed as
\bea\label{act3f}
S_{6D}&=&\int \bigg[\tr\{
-{1\over 2 } 
( \zeta_{(1)} \wedge F_{(2)}) \star( \zeta_{(1)} \wedge F_{(2)}) 
 -{1\over 2} (\zeta_{(1)} \wedge DX^I)\star ( \zeta_{(1)} \wedge DX^I) 
 \br &&
~~~~~~~ 
 +{1\over 4} (\zeta_{(1)} \wedge\star \zeta_{(1)})([X^I,X^J])^2 \}
+
\zeta_{(1)} \wedge dC_{(4)} +
\zeta_{(1)} \wedge d\theta \wedge \tr(F_{(2)}\wedge F_{(2)} )\bigg] \br
\eea
Below we write down the equations of motion 
and the constraints which follow
from  the action \eqn{act3f}.  
Note, following from the observations in \cite{hs12}, we can always 
 introduce  non-Abelian  tensors 
\bea\label{hl5s}
H_{(3)}^a\equiv \zeta_{(1)}
\wedge F_{(2)}^a +\star \zeta_{(1)} \wedge F_{(2)}^a
\eea
in the equations of motion. 
These are
 self-dual by construction, $\star H_{(3)}=H_{(3)}$. 
The index $a$ is the adjoint color index. 
By the above definition $H$ is not an independent 
tensor field, but is related to the Yang-Mills fields. 
The equations of motion can be presented as
\bea\label{eqj4} 
&& d\zeta_{(1)}=0 \br
&& DH_{(3)}^a - \star (\zeta_{(1)} \wedge DX^{Ic})X^{Ib}f^{abc}=0 \br
&& D  \star (\zeta_{(1)}\wedge DX^I)+  (\star\zeta_{(1)}) [X^J,[X^I,X^J]]=0\ 
\eea
The equation of motion for  $\eta_M$ leads to the following constraint
 \be\label{eqn39}
\tr \left(- F_{(2)}\wedge \star (\zeta_{(1)}\wedge F_{(2)}) 
+ d \theta \wedge F_{(2)}\wedge F_{(2)}\right) 
+\tr  DX^I \wedge \star (\zeta_{(1)}\wedge DX^I) 
+ \star\zeta_{(1)} V_X 
 = dC_{(4)}
\ee
where we defined  $V_X={1\over 2} \tr ([X^I,X^J])^2$. The other constraints,
some of which are  automatic consequence of the above equations, are
\bea
&& \zeta^M D_M F_{PQ}^a=0, ~~~~ 
\zeta^M F_{MQ}^a=0,~~~\zeta^M D_M X^I=0,~~~~ \zeta^M D_M H_{NPQ}^a=0
\eea
Note that in the above
eq.\eqn{eqn39} is the only equation which involves $\theta$ field explicitly. 
The rest of the equations in \eqn{eqj4} are  dependent on $\theta$ only
through $\zeta_M$. So the $\theta$ dependence in them
 can always be gauged away. (We may
wish to interpret $\tilde{F}_{(5)}\equiv dC_{(4)}$ 
as some Hodge-dual 5-form field strength.) 
However, for fully localized  e-string and I-string solitons of this theory   
we would be able to set ${C}_{(4)}=0$. Thus it is  advantageous
to have  $\theta$ dependent term in the M5 action.
The  constraint \eqn{eqn39} suggests that the local
gauge field dynamics remains largely  unaffected by the inclusion of 
$\theta$ field in the action. While
 the   $\theta$ equation of motion imposes an important
 conservation law 
\be
d\star J_{(1)}^{i-string}= 0
\ee
where the instantonic current
\be
 J_{(1)}^{i-string}= \star
{1\over 2}  \tr  ({\zeta_{(1)}}\wedge F_{(2)}\wedge F_{(2)})
\ee
The current would always be nontrivial in  presence of 
 Euclidean YM instantons.
Note, the  current is trivially conserved due to the identity
 $d\zeta_{(1)}=0$ and  the gauge Bianchi  $DF_{(2)}=0$. 

\section{5D super Yang-Mills theories}

{\it Compactification on circle with space-like $\zeta$ :}
	it is evident from  the
action \eqn{act3a} that  $\zeta^M$ is an  auxiliary field
and its equation of motion always 
requires it to take a constant value on-shell. Thus 
$\zeta^M$  can inevitably
pick  a particular spatial direction in the vacuum. As
a result the  $SO(1,5)$  symmetry of the action
gets spontaneously broken
 to  the $SO(1,4)$  subgroup in the vacuum. 
Hence, in a given vacuum the symmetries of the 6D
fields will be exactly the same  as that of 5D SYM.  In any case,
the circle compactification involves the vev $\zeta^M=|\zeta| \delta^{M}_5$, 
the radius of circle $R_5$, on which 6D theory is compactified. 
We introduce the 5D fields $(\tilde X^I,~\tilde A)$ (written with tilde 
  so as to distinguish them from 6D fields) and
relate them to their 6D counterparts as below
\be
\tilde X^I(x)=X^I(x^\mu)
\label{jku3}\ee
while gauge fields are related as
\be
\tilde A_\mu(x)=A_\mu(x^\mu), ~~~~
 A_5(x)=0.
\label{jku4}\ee
The 5D fields have no dynamics along $x^5$ (the
isometry direction) and depend only on  $x^\mu$'s. 
The axionic term and the $C_{(4)}$ term in the action \eqn{act3a} 
become total derivative and will drop out.
The action \eqn{act3a}  then reduces to the following
5D SYM action 
\bea\label{act3sym1}
S_{SYM}&=& 
{2\pi R_5 |\zeta|^2}
\int d^5x \tr \bigg[
-{1\over 4}  (\tilde F_{\mu\nu})^2  
 -{1\over 2} (D_\mu  \tilde X^I)^2 +{1\over4}( [\tilde X^I,\tilde X^J])^2
  \bigg]
\eea
Hence the 5D YM coupling constant can  be defined as \cite{hs11}
\be
{2\pi R_5 |\zeta|^2}
\equiv  {1\over g_{_{YM}}^2} .
\label{jku2}\ee
Thus the SYM action one gets is
\bea\label{act3sym}
S_{SYM}&=& 
{1\over g_{_{YM}}^2}
\int d^5x \tr \bigg[
-{1\over 4}  (F_{\mu\nu})^2  
 -{1\over 2} (D_\mu  X^I)^2 +{1\over4}( [X^I,X^J])^2
  \bigg]
\eea
where tilde sign over 5D fields has been dropped. 

Note that $|\zeta|$ does not explicitly appear in 5D action
and thus could remain arbitrary. 
Let us add a few clarifications here. 
The $|\zeta|$ has  dimensions of mass. 
Upon compactification another 
scale available in the theory is the radius of compactification $R_5$. 
So we can at best claim  that
 \be
|\zeta|^2 = {1\over 2\pi R_5}  {1\over g_{_{YM}}^2} .
\ee
However, using the fact that 
5D SYM theory has only one dimensionful parameter in the form of  
 coupling constant $g_{_{YM}}$, and the expected relation \cite{doug10}
\be
  2\pi R_5\simeq {g_{_{YM}}^{2} }, 
\label{jku2a}\ee
we  find it appropriate to fix
\be
|\zeta|\simeq{1\over g_{_{YM}}^2}.
\label{jki2a}\ee
Also we shall see in this paper that this remains a consistent choice. 

Before we close this section, let us also mention that
{\it a  time-like reduction} of the action \eqn{act3a} 
can also be performed, where we will  take
$\zeta^M=({1\over g_{_{YM}}^2},0,0,0,0,0)$,  a  time-like vector.
This will lead to  5-dimensional  Euclidean theory 
\bea\label{act3sym2}
S^E&=& 
{\beta\over g_{_{YM}}^4}
\int d^5x \tr \bigg[
{1\over 4}  (F_{\mu\nu})^2  +
 {1\over 2} (D_\mu  X^I)^2 -{1\over4}( [X^I,X^J])^2
  \bigg]
\eea
where $\beta\sim O( g_{_{YM}}^2)$ is a free parameter with a dimension of length. 
Note that the
signs of all the terms have become
 opposite of the Lorentzian SYM case \eqn{act3sym}.

\section{Instantonic string  solitons}

We now show that there   
exist  instantonic  solutions in the 6D theory
with a completely new interpretation!
These are extended  string like solitons  which  live on
  M5-brane stack and will carry instantonic charges. 
Consider an ordinary (e)lectrically charged
string soliton on M5-brane, being stretched, say, 
along  $x^5$  direction.
These infinitely extended  e-strings  do not have
momentum  along the  direction of  extension (isometry). Thus
${\bf p}_5=0$  typically for all ordinary e-string  states. But 
for the I-string solutions we present here
\be
{\bf p}_5 \propto {\rm instanton~number} 
\ee
and it will be nonvanishing and also quantized.

In the previous version of the theory \cite{hs11}, 
the instantonic solutions could be found, but that required the presence of 
non-zero flux  for $C$-field. In the present avatar of the theory,  
with the inclusion of a $\theta$-term in the 
action \eqn{act3a}, the $C$ flux  can be set to zero.
In the purely instantonic sector with vanishing 
$C$, and taking all $X^I$'s constant 
diagonal matrices, the  constraint  equation \eqn{eqn39} simply
reduces to the condition
 \be\label{eqn45}
  \star \zeta_{(1)}\wedge F_{(2)} 
- d\theta \wedge F_{(2)} = 0
\ee
Now we consider  the gauge instanton configuration 
$\star_4 F_{(2)}=F_{(2)}$ localised in 4D Euclidean subspace
$E_4$ of the 6-dimensional spacetime 
${\cal M}_{2}\times  E_4$, whereas
both the $\zeta_M$ and the $\theta$  are confined to live  in 2D Minkowski
space, ${\cal M}_2$.  Then the  
condition \eqn{eqn45} will be satisfied if 
\be\label{eqn46}
 \star_2 \zeta_{(1)} =   d \theta\ .
\ee
That can be solved provided  
\be 
\eta_M=0,~\zeta_M=\partial_M \theta \ ,
\ee
in which case $\theta$ will have to be    
\be \label{eqn46t}
\theta(x^+,x^-)=g x^+ \ .
\ee
Hence $\zeta_{(1)}$ is a chiral (self-dual) 1-form
in 2D  flat spacetime, i.e.  $\zeta_M=
|\zeta| \delta_{M+}=g \delta_{M+}$, $(\zeta_M)^2=0$. 

The above instantonic soliton  describes a nontrivial string like configuration.
The $\theta$ equation of motion
simply implies a conserved `instantonic' current
\be
 J_{i-string}= \star
{1\over 2}  \tr  ({\zeta_{(1)}}\wedge F_{(2)}\wedge F_{(2)})
\ee
This current will be nontrivial  whenever gauge
instantons are present.  It will also be light like, so
\be
J^M_{i-string}=j(x)(1,0,0,0,0,1). \ee 
The instanton charge carried by the I-string  becomes
\be
Q_{i-string}=\int d^5x J_{i-string}^{0} =
{1\over 2}
\tr \int_{\Sigma_5} {\zeta_{(1)}} \wedge F_{(2)}\wedge F_{(2)} \ .
\ee
It  is a dimensionless quantity in 6D.  
The distribution of the instanton charge however  depends on  $\zeta_M$
vev in a given  vacuum. Following from \eqn{eqn46}
and \eqn{eqn46t}, let us take  the vev
$<\zeta_M> =|\zeta|(1, 0,0,0,0,1) $. 
In this vacuum the instantonic string is
localized in $x_1,\cdots, x_4$ (4-dimensional Euclidean space  $E_4$), 
and the instanton charge gets uniformally spread out along remaining  
flat direction $x^5$. This may lead to an infinite
result as the charge is being  carried by an extended
string like object. 
However we can define the charge per unit length  of  I-string 
(taking length $L \equiv \int dx^5$ as a regulator) 
\be\label{j8k}
\rho_{i-string}={Q_{i-string}\over L} = {|\zeta|\over 4}
\tr \int d^4x F_{mn} F^{mn}=n |\zeta| 
\ee
 which is certainly quantized and finite. The integer  $n$
is  the count of the  instanton number. Thus from eq.\eqn{j8k}
we get an independent
interpretation for the $\zeta$ vev.
That the parameter $|\zeta|$  is  the
measure of  I-string charge density in a given M5 vacua.
While $L$ is simply the regulator of the I-string length. 
The  quantities, $L$ and $\zeta$, are so far quite independent. But
when we consider circle compactification, since $x_5\sim x_5+2\pi R_5$, 
we may identify compactification radius
\be
2\pi R_5  \simeq {1\over |\zeta| }\simeq
g_{_{YM}}^2\ ,
\ee
as argued previously, because 
 5D SYM theory has only one dimensionful parameter, namely
$g_{_{YM}}$). Thus we   conclude that the
 6D I-string solitons are  made up of the gauge  instantons,
and the  $|\zeta|$  provides  an independent  scale to
 measure  their charge densities. However
 abelian  theory for a single M5-brane  cannot apriori  admit
I-string vacua. 

\subsection{The energy-momentum tensor}  
The energy-momentum tensor 
can be found from the new action \eqn{act3a}. It is given by 
\bea
&& T_{MN}=\tr\bigg[
{1\over 2.2!}(  G_{MPQ} G_N^{~PQ} -{\eta_{MN}\over 6} G_{PQR}G^{PQR})+
{1\over 2}(  C_{MP}^{I} C_N^{I,P} -{\eta_{MN}\over 4} C_{PQ}^{I} C^{I,PQ})\br
&&~~~~~~~~- ({\rm potential ~terms})
\bigg]
 \eea
where we have denoted 
$G_{MPQ}\equiv \zeta_{[M}F_{PQ]}$ and  
$C_{MP}^I\equiv\zeta_{[M}D_{P]}X^I$ for simplificity. The  details 
of the potential terms are not important as the contribution 
from these terms will be vanishing for the I-strings.
For the I-string vacua we get the energy
\bea
E= \int d^4x [dx^5] T_{00}&=& {L\over 4}
\int d^4x 
\zeta_0\zeta_0   \tr F_{mn}F^{mn} \br
&=&  n L |\zeta|^2  \  .
\eea
Thus the energy per unit length of the I-string states is
\be
{E\over L}= n 
|\zeta|^2  \ge 0
\ee
Similarly, the 5-th component
\be
{\bf p}_5= \int d^4x [dx^5] T_{05}= 
{L\over 4}\int d^4x 
\zeta_0\zeta_5 \tr F_{mn}F^{mn}
=  n L {|\zeta|^2}  
  .
\ee
(For the non-instantonic solutions, however, ${\bf p}_5=0$.) 
Other components such as $T_{0m}$, for $(m=1,2,3,4)$ are all vanishing except
\be
T_{55}=  {1\over 4} |\zeta|^2 \tr F_{mn}F^{mn}.
\ee
For the   I-strings  we can calculate 
\be\label{jk865}
M^2=-p_Mp^M=E^2-p_5^2=0\ .
\ee
It  implies that the  I-strings 
behave like exactly massless solitons, with a lightlike $\zeta_M$  (non-zero components 
$\zeta_0= \zeta_5=|\zeta|$),
and  these  states can be described by the momentum vector
\be
P^M=(
{nL|\zeta|^2},0,0,0,0, {nL|\zeta|^2})
\ee
Thus the I-strings are exactly lightlike states in 6D with discrete momenta.
 
As described above, the I-string  solutions
are specifically obtained when we take $\zeta_M$ 
having components along the lightcone,    
$\zeta_M=|\zeta| \delta_{M+}$, the axion field $\theta=|\zeta| x^+$, while
  $F$ is taken to be the Yang-Mills  self-dual 2-form
living over the patch $(x^1,x^2,x^3,x^4)$. Accordingly
the self-dual 3-rank tensor $H$  can be constructed
\be 
 H_{(3)}=|\zeta| dx^+\wedge (F_{(2)} +  \star_4 F_{(2)}),
\ee
 it satisfies $dH_{(3)}=0=d\star H_{(3)}$, also see previous definition
in \cite{ hs12}.

\subsection{ Light-like  $\zeta^M$  Vs strong SYM  coupling}

We consider $\zeta^M=|\zeta|(1,0,0,0,0,1)$ 
being a constant   vector field. With $X^I$'s being
 constant  diagonal matrices, the
potential  for  adjoint scalars also vanishes. 
In such cases the 6D bosonic  action will 
 mainly contain the Yang-Mills fields
\bea\label{act3ae}
S_{instanton}&\simeq&\int d^6x \tr \bigg[
-{1\over 12 } ( \zeta_{[M}F_{NP]})^2  
 +{1\over 4}\epsilon^{MNPQRS}
\zeta_M\partial_N \theta F_{PQ}F_{RS} \bigg]
\eea
We can see that for the I-string solutions, 
supported by lightlike $\zeta$ and $\theta$, this action 
 vanishes on-shell. This tells us that the action supports
  lightlike configurations and is meaningful.
Next let us take $x^5$  to be  compact, $x^5\sim x^5+2\pi R_5$,  
we also set $L \sim 2\pi R_5$, and identify  
\be
 L |\zeta|^2 \simeq {1\over g_{_{YM}}^2}.\ee
 Then the limit $|\zeta| \to 0$   corresponds to
 5D coupling constant $g_{_{YM}}\to \infty$. 
Thus in  this limit the 6D theory on a circle can
represent a truely strongly coupled phase of 5D SYM. 
But we need to make sure that this limit keeps the
I-string charges finite; 
\be
 Q_{i-string}= Z_{monopole} L |\zeta|  = {\rm Finite} 
\ee
whereas the monopole (soliton) charge  in 5D is given by
\bea
Z_{monopole} = {1\over 2 }\tr \int   F\wedge F
={n}\ .
\eea
Although the  instantons are
strings from 6D  perspective but from 5D 
point of view they are  the  monoples 
with fixed topological charge $Z$. 
The  monopole   mass in 5D is given by
\be
M_{monopole}= { n\over  g_{YM}^2}.
\ee 
Note for the 6D I-strings, as we found $M_{i-string}^2=0$,  instead
we have
\be
P^M=({ n\over  g_{YM}^2}
,0,0,0,0, { n\over  g_{YM}^2})
\ee
Thus the massive monopoles are lifted to 
the I-strings which form a discrete spectrum of lightlike states in 6D. 
Hence we can  describe  strong coupling limit, 
  $g_{_{YM}}\to\infty$ and $M_{monopole}\to 0$, as a limit under which all
discrete I-string states with distinct topological charges
becoming degenerate   and   condensing
 to  the ground state in 6D theory. In order to  keep
$Q_{i-string}$ finite,  following large box limit has 
to be employed
 \be
  L=2\pi R_5\to \infty , ~~~|\zeta|={1\over g_{_{YM}}^2}\to 0,~~ L|\zeta|={\rm fixed}\ .
\ee
Thus,  strong coupling limit  corresponds to infinitely long 
I-strings 
 condensing to the ground state in a  decompactified     
M5 theory \eqn{act3a}. Although being degenerate in the ground state, each
I-string vacua  carries distinct  topological charge 
 $Q_{i-string}$. The charge density
$\rho_{i-string}\to 0$,
as the  topological charge gets spread  uniformly along  infinitely
 long (tensionless) I-string.

\section{ e-strings and self-dual $H$}
Let us  note that all  6D solutions in our theory will have at least 
one isometry direction due to the nontrivial constant vev of the dressing 
field $\zeta_M$ \cite{hs11}. 
It is also evident from the construction of the 6D action
that there will be no  point-like localized solutions.
 The  supersymmetric e-string vacua in  6D Abelian chiral theory
 have been described in
\cite{howe1}. We  study these solitonic
configurations describing the intersection of an  
extended M2-brane ending  on M5-brane.
Consider the $\zeta$ vacuum where  $\zeta^M=(0,0,0,0,0,|\zeta|)$, 
   aligned along $x^5$,
 which we take to be the isometry direction. 
 Note $|\zeta|$ has dimensions of mass and it is 
the only dimensionful parameter available in the vacuum.
This solution is described by \cite{hs11}
\bea\label{hl2} 
&& X^I(x)= \delta^{I 10} \Phi(x_m),~~~~(I=6,7,8,9,10) \br
&& F_{0m}=\pm  \partial_m \Phi .
\eea
It  is a solution of equations \eqn{eqj4} and \eqn{eqn39} provided
\be 
\Phi(x_m)=\Phi_0+{ q\over |\zeta| (x-x_o)^2}
\ee
The fields are dependent only on  $x_m$ ($m=1,2,3,4$) coordinates, and
 $\Phi_0$ is an arbitrary dimensionful constant signifying 
the asymptotic value of the field. 
This solution can be easily generalized to  multi-centered solutions,
 where $\vec{x}_{o (i)}, ~q_{(i)}$,  index $(i=1,2,\cdots,r)$, 
will parametrize respective 
positions and charge of  $i$-th soliton. 
Since  only one  scalar, namely $X^{10}$ 
(representing the coordinate $x^{10}$ transverse to M5),
 has been excited, it
describs an  M2-brane, in the $x^5$-$x^{10}$ plane, ending  
on  M5-brane. The string 
is  along the common  intersection direction of the branes, which is  $x^5$. 
The electric field surrounding the string defect is
peaked at its location $\vec x=\vec x_o$ and depends upon the vev $|\zeta|$. 
Thus for  e-strings solution to exist we need to have nontrivial $|\zeta|$.
We also determine the nonvanishing components 
of the $H$ tensor in the equations \eqn{hl5s}
\be
H_{0m5}=\zeta_5 F_{0m}=\pm |\zeta| \partial_m\Phi, ~~~~
H_{mnp}= \epsilon_{mnp0l5} \zeta^5 F^{0l}=
\pm |\zeta| \epsilon_{mnpl}\partial^l\Phi
\ee
where we took $\epsilon^{012345}=-\epsilon_{012345}=1$, and 
$\epsilon_{1234}=1$ is the Levi-Civita tensor in four Euclidean
dimensions. It shows that the $H$   is self-dual. 
Remarkably, no   background value for $C$ tensor  
is required for the e-string solutions, this is because  the constraint
\eqn{eqn39} reduces  to
 \be\label{eqn39a}
 F_{(2)}
\wedge \star (\zeta_{(1)}\wedge F_{(2)}) 
-  DX^{10}\wedge \star (\zeta_{(1)}\wedge DX^{10})=0 
\ee
which is automatically satisfied since $F_{0m}=
\pm \nabla_m X^{10}=\pm \partial_m \Phi$.

The charge  carried by   e-strings  is given by
\bea
&& Q_{e-string} ={1\over 4\pi^2}\int \star H_{(3)}
={1\over 4\pi^2}\int_{ S^3}  H_{(3)}=  q \ .
\eea
We must also note that 
\be
Q_{electric}=Q_{magnetic}
\ee because these are  self-dual strings. By
 taking the regulated length of the e-string as $L$, 
we define  e-string charge density 
\bea
&&\rho_{e-string} \equiv {Q_{e-string} \over L}=q |\zeta|  
\eea
where $q$  is a measure of the net  charge carried by self-dual string.

{\it Compactification and the vanishing $|\zeta|$  limit}:
in order to connect  e-strings with the 5D SYM states, 
we need to compactify them 
on a circle, with the given prescription
 \be
L=2\pi R_5\simeq g_{_{YM}}^{2}, ~~~~ |\zeta|=g_{_{YM}}^{-2} \ee
as discussed earlier.
In the strong coupling regime  the e-string 
charge density will become vanishingly small, 
but the net charge $Q_{e-string}$   stays finite. 
Remarkably the $H$ 
field exists even in the strong coupling limit 
(or vanishing $|\zeta|$ limit)
\be
{\rm ~Lim}_{|\zeta|\to 0} ~H_{0r5} = {2q\over r^3}. 
 \ee
The energy density of the e-string soliton is
\bea
&& p_0 \simeq
L|\zeta|^2\int d^4x (\vec E)^2 =
 L \int  d^4x  [{2q^2\over r^6}] \br
&&~~~\sim g_{_{YM}}^{2}\int d^4x 
{\cal E} 
\eea
where ${\cal E} \equiv {2q^2\over r^6}$ is the measure of 
corresponding energy flux in 5D SYM.
The energy of the elemetary  e-string states 
grows linearly with  $L$  (or $ g_{_{YM}}^{2}$). Hence
the e-strings tend to become  heavy as YM coupling becomes large.
Note these were the lightest states in the perturbative regime in SYM,
while  monopoles were
 heavy, as the monopole mass is ${n\over g_{_{YM}}^2}$. 

\subsection{ Dyonic solutions: $F_{0i}\ne 0$ and $F_{ij}\ne 0$}
The  dyons  have both electric and magnetic $F_{MN}$
components and generally occur as localized solutions 
in 4D SYM. For embedding these in 6D, we need to consider $x^4$ and 
$x^5$ both to be the isometry directions. 
The constant $\zeta$ configuration is 
\bea\label{hl2s} 
&& \zeta^M=(0,0,0,0,0,|\zeta|),  ~~~~C_{(4)}=0, \br
&& X^I= \delta^{I 10} f(x), ~~~~
F_{0i}=\pm  \partial_i f_e,  
 ~~~~(i,j=1,2,3) (I=6,7,8,9,10)
\eea
This electrically charged $(q_e,0)$ solution is  localized over
three  Cartesian coordinates $(x^1,x^2,x^3)$ only,
and is a solution of equations \eqn{eqj4}  and constraint \eqn{eqn39} with
\be 
f(x)=
f_e(x)=f_0 +{ q_e \over  |x-x_o|}, ~~~~
\ee
where $f_0$ is a dimensionful constant. At
this point this  purely is a result of the constraint \eqn{eqn39} present 
in the  theory. Usually this follows from the supersymmetry.

The magnetically charged solutions with
$(0,q_m)$  are not  covered by the above ans\"atze.
Furthermore, the
action of  electric-magnetic $SL(2,Z)$  transformations
on a $(q_e,q_m)$ dyon will certainly lead to a jump in the value 
of $q$'s. \footnote{ The  SL(2,Z)-duality transformation of the dyon charges: 
$$
\left( \begin{array}{c}  q_e \\  q_m   \end{array} \right) \to 
\left(  \begin{array}{cc}  a & b \\  c & d   \end{array} \right) 
 \left( \begin{array}{c}  q_e \\  q_m   \end{array} \right)
$$  
where integers obey the condition
 $ a d - b c = 1 $. } 
But, in order to obtain a magnetically charged solution  we 
need to switch on   $C_{(4)}$ background. 
A  magnetic $(0,q_m)$ solution is found with following ansatz 
\bea\label{hl2sm} 
&& \zeta^M=(0,0,0,0,|\zeta|,0), ~~~~~
\eta_M=0,~~~C_{05\theta\phi}={2 |\zeta|q_m^2\over  r}, \br
&& X^I= \delta^{I 10} f(x), ~~~
F_{0i}=0, ~~~ 
F_{ij}= \epsilon_{ijk} \partial_k f, ~~~~(i,j=1,2,3) 
\eea
where function $f(x)=f_0 +{ q_m \over  |x-x_o|}$. 
The  $(r,\theta,\phi)$, 
 with $r^2= (x^i-x_o^i)^2$, 
are the spherical coordinates  around the center ${\vec x}_o$. 

Thus we  may understand that  strong-weak 
duality of 4D SYM acts in  nontrivial way on the 6D fields, 
including  auxiliary fields.
Let us  make it more explicit. 
We separate  6D coordinates as $x^M=(x^\mu,x^m)$, 
with greek indices $\mu,\nu=0,1,2,3$ and small roman indices $m,n=4,5$.
We choose a constant 6D vector $\zeta_M=(0,0,0,0,\zeta_4, \zeta_5)$. We shall
assume that no background field depends upon $(x^4,x^5)$ except $\theta$
(taking $\eta_M=0$).
 One can introduce dual pairs: 
$\xi_m=(\zeta_4, \zeta_5)$ and dual $\tilde \xi_m=(\zeta_5, -\zeta_4)$. 
Both are  2-vectors in the $x^4-x^5$  plane.
Similarly given $C_{4\mu\nu\lambda}$ and $C_{5\mu\nu\lambda}$, being only
nonvanishing components, we can
 define 
$C_{m\mu\nu\lambda}=(C_{4\mu\nu\lambda},~C_{5\mu\nu\lambda})$ 
and corresponding
 dual components
$\tilde C_{m\mu\nu\lambda}=(C_{5\mu\nu\lambda},~-C_{4\mu\nu\lambda})$. 
From these  we  construct  following $SL(2,Z)$ doublets: 
\bea
&&\xi^{a}_m=(\tilde \xi_m ,  - \xi_m) , ~~~~ (a=1,2) \br
&&{\cal F}_{\mu\nu}^a= ( F_{\mu\nu}, ~\tilde{F}_{\mu\nu}) \ .
\eea 
We can now rewrite $H$-tensor (nonvanishing components only) as
\be
H_{m\mu\nu}=\xi^a_m {\cal F}_{\mu\nu}^b\omega_{ab} 
=
\left(\ba{cc} \tilde \xi_m &  - \xi_m \ea \right)
\left(\ba{cc} 0 & 1\\ -1&0  \ea \right)
\left( \ba{c} F_{\mu\nu} \\ \tilde F_{\mu\nu} \ea\right) 
\ee
which is an $SL(2,Z)$  invariant expression, where 
$$\omega_{ab}=\left(\ba{cc} 0 & 1\\ -1&0  \ea \right)
$$
is  $SL(2,Z)$ metric.
The constraint equation \eqn{eqn39} can also be rewritten as
\bea
{1\over 2}{ \cal F}^{\mu\nu a} H_{m\mu\nu} 
-\omega_{ab}\xi_m^b (\partial_\mu X^{10})^2
=  {\cal G}^{ a}_m 
\eea
where the $SL(2,Z)$ doublet on  right hand side is
\bea\label{gc5}
{\cal G}^{a}_m 
\equiv \left( \ba{c} 
  {1\over 3!}\epsilon^{\mu\nu\lambda\rho}\partial_{\mu} C_{\nu\lambda\rho m}
+{\tilde \zeta}_{m} F_{\mu\nu} \tilde F^{\mu\nu} \\
 -  {1\over 3!}\epsilon^{\mu\nu\lambda\rho}\partial_{\mu}{\tilde C}_{\nu\lambda\rho m}
- \zeta_{m} F_{\mu\nu} F^{\mu\nu} \ea \right)
\eea
These define the covariance of  6D  equations 
under $SL(2,Z)$ group.
Thus in dyonic cases $C_{(4)}$ backgrounds can be nontrivial, as we have 
seen it already in the magnetically charged solution  \eqn{hl2sm}. 
Also it appears that for a given dyonic solution,
one could always set a gauge choice  such that, 
for both $C$  and $\tilde C$ in \eqn{gc5}, we get
${\cal G}^{a}_m=0$. But this choice may altogether
be guided by (2,0) supersymmetry in the complete theory, which we have not 
explored in the present work.

\section{Conclusion}

In summary,
the covariant 6D gauge action  presented here
with an auxiliary vector, $\zeta^M$, and   
an associated axion, $\theta$, 
can effectively describe the low energy dynamics on M5 branes. 
The theory admits  namely two types of extended solitonic solutions.
For the I-string  solitons, living over ${\cal M}_2\times E_4$ spacetime, 
with  lightlike vev 
$\zeta^M=(|\zeta|,0,0,0,0, |\zeta|)$, 
the  mometum is  also  
lightlike  
$$p^M=(n {L|\zeta|^2},0,0,0,0, n {L|\zeta|^2}).$$ 
The  mometum  effectively depends on  the instanton  
number $n$ carried by the I-string soliton. 
The I-strings, described as strings of length $L$, are a set of 
discrete states which
carry a  topological charge 
$$
Q_{i-string}=n L|\zeta|\ .$$
When theory is compactified on a circle,
 taking ${L|\zeta|^2}\sim {1\over g_{_{YM}}^2}$, 
the 6-momentum carried by I-string becomes 
$$p^M=( {n\over  g_{_{YM}}^2},0,0,0,0,  {n\over  g_{_{YM}}^2}) $$
but it is light-like always. Thus
we emphasize that the  I-strings  arise only in the lightlike case
 $(\zeta)^2= 0$ and the axion is chiral field. 
The 6D on-shell action is  well defined. 

For the ordinary (color) e-string solitons, 
with  space-like   $\zeta^M=(0,0,0,0,0, |\zeta|)$, 
 the symmetry  allows  the 6-momentum to be
$p^M\propto (L {\cal E}, 0,0,0,0,0)$
which interestingly remains independent of the $\zeta$-vev. 
The charge  of   e-strings  is given as
\bea
&& Q_{e-string} =\int \star H_{(3)}
=\int  H_{(3)} .
\eea
While  full $SO(1,5)$ invariant 
configurations  in the theory  can occur 
only when vev $<\zeta_M>$ vanishes. Also we  cannot implement 
vanishing $|\zeta|$ limit arbitrarily because all 6D string  
configurations are defined in a regulated space with the
string length $L$. However, the limit $|\zeta|\to 0$
 keeping $L|\zeta|={\rm fixed}$ can be smoothly implemented, 
which  keeps the  i(e)-string charge fixed. 
In the theory compactified  on a circle,
 this double limit  coincides with the strong SYM coupling limit, 
$L\sim g_{_{YM}}^2\to\infty$,
$|\zeta|\sim {1\over g_{_{YM}}^2}\to 0$,
 of 5D  super Yang-Mills theory. The strong coupling 
limit in SYM is usually  thought of as  
the decompactification limit from 6D point of view. In
the strong coupling limit the infinitely long
I-strings will become very light (tensionless)  and 
become degenerate with the 6D ground state,
while  colored e-strings  tend to become  heavy because
 $L\sim g_{_{YM}}^2\to\infty$.

Thus we  conclude that the vacuum in M5 theory would be infinitely degenerate 
and will be proliferated by multiple (tensionless) I-strings. Each such
I-string vacua would carry a distinct topological (instatonic) charge. 
The degenerary gets lifted
as soon as the M5 theory is put on a finite size circle.

\vskip.5cm
\noindent {\it Acknowledgments:}

 I am  grateful to  D. Sorokin for helpful discussion and the 
 careful comments on the draft. I  thankfully 
acknowledge the associateship support
from the ICTP, Trieste.

\end{document}